\documentclass[useAMS,usenatbib,referee]{mn2e}

 \title[IPHAS-POSS-I Proper Motion Survey]{The IPHAS-POSS-I proper motion survey of the Galactic Plane}
 \author[N.R.\ Deacon et al.]{N.R.\ Deacon\thanks{E-mail:
 ndeacon@astro.ru.nl}$^1$, P.J. Groot$^1$, J.E. Drew$^2$, R. Greimel$^{3,4}$, N.C.\ Hambly$^5$, M.J. Irwin$^6$,\newauthor A. Aungwerojwit$^{7,8}$, J. Drake$^9$, D. Steeghs$^{8,9}$, \\
$^1$Department of Astrophysics, IMAPP, Radboud University Nijmegen, P.O. Box 9010, 6500 GL Nijmegen, The Netherlands\\
$^2$Centre for Astrophysics Research, University of Hertfordshire, College Lane, Hatfield AL10 9AB\\
$^3$Isaac Newton Group of Telescopes, Apartado de correos 321, E38700 Santa Cruz de La Palma, Tenerife, Spain\\
$^4$Institut f\"{u}r Physik, Karl-Franzen Universit\"{a}t Graz, Universit\"{a}tsplatz 5, 8010 Graz, Austria\\
$^5$SUPA\thanks{Scottish Universities' Physics Alliance}, Institute for Astronomy, 
School of Physics, University of Edinburgh, Royal Observatory Edinburgh,\\Blackford Hill, Edinburgh, EH9 3HJ\\
$^6$Institute of Astronomy, Madingley Road, Cambridge CB3 0HA\\
$^7$ Department of Physics, Faculty of Science, Naresuan University, Phitsanulok, 65000, Thailand\\
$^8$ Department of Physics, University of Warwick, Coventry, CV4 7AL, UK\\
$^9$ Harvard-Smithsonian Center for Astrophysics, Cambridge, MA 02138, USA\\
}
 \begin{document}
 \date{}
 \pagerange{\pageref{firstpage}--\pageref{lastpage}} \pubyear{2005}
 \maketitle
 \label{firstpage}
 \begin{abstract}
We present a proper motion survey of the Galactic plane, using IPHAS data and
POSS-I Schmidt plate data as a first epoch, that probes down to proper
motions below 50 milliarcseconds per year. The IPHAS survey covers the
northern plane ($|b| < 5^{\circ}$) with CCD photometry in the $r$, $i$ and H${\alpha}$
passbands.  We examine roughly 1400 sq. deg. of the IPHAS survey area and draw up a catalogue containing 103058
objects with significant proper motions below 150 millarcseconds per year in the magnitude range 13.5$< r' <$19.  Our survey sample
contains large samples of white dwarfs and subdwarfs which can be identified
using a reduced proper motion diagram. We also found several objects with IPHAS
colours suggesting H${\alpha}$  emission and significant proper motions. One is the known cataclysmic variable GD552; two are known DB white dwarfs and five others are found to be non-DA (DB and DC) white dwarfs, which were included in the H$\alpha$ emission line catalogue
due to their lack of absorption in the H$\alpha$ narrow-band.

 \end{abstract}
 \section{Introduction} 
The INT Photometric H$\alpha$ Survey (IPHAS, Drew et al., 2005) is a deep
($r<$21), CCD based survey in three filters ($r$, $i$, H$_{\alpha}$) covering
1800 sq. deg. of the northern Galactic Plane ($|b|<5^{\circ}$). IPHAS forms part of the European Galactic Plane Surveys (EGAPS), which also includes the UKIRT Infrared Deep Sky Survey
(UKIDSS) Galactic Plane Survey (Lawrence et al., 2007, Lucas et al. 2008) covering 1800 sq. deg. of
the plane in $J$, $H$, $K$ to a depth of $K$=19 and the UV EXcess survey (UVEX, Groot et
al. in prep.). UVEX is planned to 
complement IPHAS by covering the same area but in $u$, $g$ and HeI 5875\AA with an
additional $r$ band epoch. These surveys also have upcoming southern
counterparts.  With the number density of stars highly
concentrated on the Plane, IPHAS and EGAPS provide ideal tools to study a whole
range of stellar and Galactic research topics. They have already yielded significant discoveries in
fields such as cataclysmic variables (Witham et al., 2007), planetary nebulae
(Mampaso et al., 2006, Wesson et al., 2008), young low mass objects (Valdivielso et al., 2009), star forming regions (Vink et al., 2008)
and extinction in the Galactic plane (Sale et
al., 2009).  Large scale CCD-based astronomical surveys such as IPHAS provide
accurate photometric and astrometric data on large numbers of astronomical
objects. In addition to their main science goals, surveys such as EGAPS make their data public (see Gonzalez-Solares et al., 2008 for details on public IPHAS
data) and they can be used by anyone in the astronomical community to pursue their own research aims. Combining IPHAS data with those from other surveys with
different wavebands or epochs can lead to discoveries of variable objects and
can also allow the parameter space of each object to be expanded to include
not only magnitudes and positions but proper motions as well. Here we
undertake the first comprehensive, optical, wide field survey to identify proper motions below 0.1 arcseconds per year
in the Galactic Plane by cross-referencing the IPHAS database with SuperCOSMOS
(Hambly et al., 2001)
scans of the POSS-I plates taken in the 1950s. This gives us a proper motion baseline of approximately fifty years.

Early proper motion surveys utilised blink comparators and exceptional
patience to indentify moving stars manually. The early manual work of Luyten
is brought together in two samples, the Luyten Half-Arcsecond Survey (LHS,
Luyten, 1979a) which catalogued 3561 objects with proper motions greater than
half an arcsecond per year and the New Luyten Two Tenths Catalogue (NLTT,
Luyten, 1979b) which contained 58845 objects with $\mu >$0.2 arcseconds per year(''/yr). Both these surveys ran into difficulties in the Galactic Plane with the NLTT survey less than 50\% complete for $|b|<15^{\circ}$ at magnitudes fainter than $V$=16 (Lepine \& Shara, 2005). The main modern computational study is that of Lepine (2008). They used a sophisticated algorithm to degrade POSS-II images to the same quality as the older POSS-I images. The two could then be subtracted and high proper motion stars identified. This survey is complete to V=20 and $\mu$=0.15''/yr. However the survey suffers from crowding in the Galactic Plane leading to a reduction in completeness. Lepine \& Shara estimate they are only 80-90\% complete down to V=19 within 15 degrees of the Galactic Plane. Fedorov et al. (2009) predict their upcoming catalogue will cover low proper motions in the Galactic Plane but will not provide a consistent proper motion range due to a varying maximum proper motion. Gould \& Kollmeier (2004) used data from the Sloan Digital Sky Survey photographic plate data to produce a proper motion survey below 100 milliarcseconds per year. However this avoided the Galactic plane. The study of Folkes et al. (2007) attempts to fill in the Galactic Plane gap left by southern surveys such as Deacon \& Hambly (2007), Pokorny et al. (2004) and Finch et al. (2007) (all of which avoid the Plane) by combining UKST and 2MASS data in a similar manner to Deacon \& Hambly (2007) to identify candidate low mass stars and brown dwarfs from their proper motion.

\section{Method}
In order to plan our proper motion survey we had to first consider the
datasets available. Two datasets are available for use as a first epoch, both having
been scanned using the SuperCOSMOS plate scanning machine (Hambly et
al. 2001). As well as the POSS-I plates, the newer, higher quality POSS-II
plates with better emulsion sensitivity and improved resolution are also available. These provide better astrometric accuracy but a
much shorter time baseline with respect to IPHAS (10-15 years compared to the IPHAS data versus the roughly 50 year epoch difference betweeen IPHAS and POSS-I). However
a shorter baseline means less contamination due to spurious pairings;
$n_{spurious}\propto (\mu_{max} \Delta t)^2 $, where $n_{spurious}$ is the
number of spurious pairings, $\mu_{max}$ is the maximum proper motion and
$\Delta t$ is the epoch difference. $n_{spurious}$ is also proportional to the
density of objects around the target. This is one of the reasons most proper motion
surveys have avoided higher density areas of the sky such as the Galactic
Plane. 
 \begin{figure}
 \setlength{\unitlength}{1mm}
 \begin{picture}(75,120)
 \includegraphics{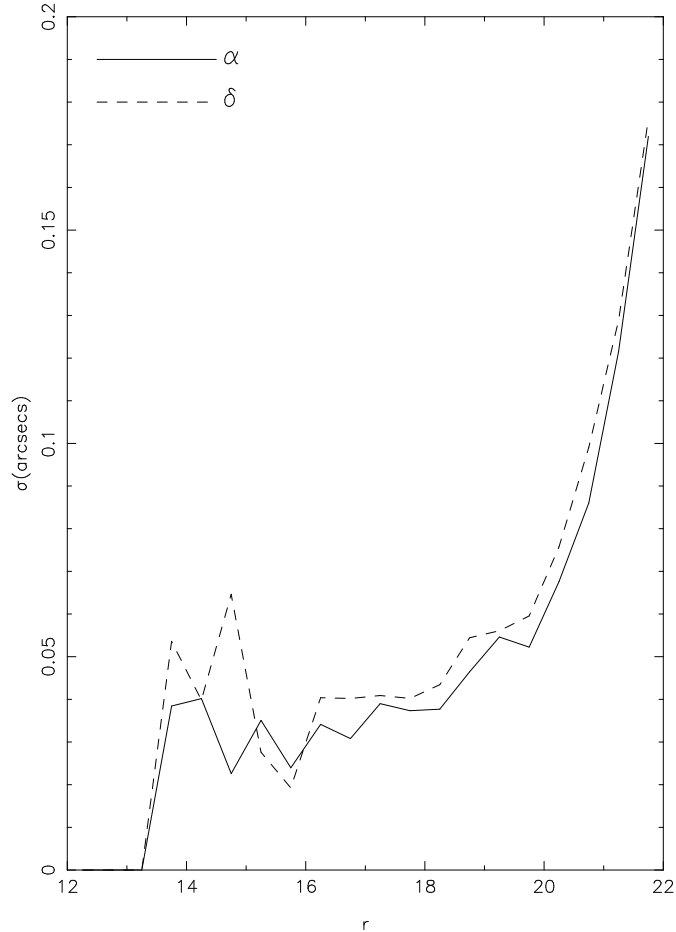}
 \end{picture}
 \caption[]{The astrometric errors (in arcseconds) between the IPHAS and UVEX surveys.}
 \label{uvexErrplot}
 \end{figure}
Along with these data we also have the upcoming UV EXcess (UVEX) survey (Groot et al., in prep.)
which will be a blue companion to IPHAS and a second $r$ epoch. This will provide us with CCD quality second
epoch astrometry, observed by the same telescope and camera, reduced by the
same pipeline but with only a 3-5 year baseline. Examining the positional
errors between IPHAS and UVEX we found that they were typically 40
milliarcseconds (see Figure~\ref{uvexErrplot}), rising to 50 mas at $r$=19 (where POSS-I plate astrometry becomes difficult, see Figure~\ref{Errplot}) and to roughly 100 mas, as the survey limit (r$\sim$21) is approached. Hence
we can assume that with a three year baseline, the minimum 5$\sigma$ proper
motion detectable between IPHAS and UVEX at the survey limit (r$\sim$21) is roughly 166 mas/yr. At the limit
at which astrometry on the POSS-I plates becomes difficult ($R_F$=19) the
minimum proper motion becomes 100 mas/yr. Hence below
this latter limit (also below the $\mu$=0.15''/yr lower proper motion limit of Lepine
\& Shara, 2008) there is the potential for a lower proper motion survey to probe to previously unexplored proper motions in the Galactic Plane. To coincide with the lower
limit of a potential IPHAS-UVEX proper motion survey (less then 100 mas/yr for objects brighter the $r$=19) and leaving some
overlap we decided on a maximum proper motion of 0.15''/yr. This means that
even with the exceptionally long baseline between IPHAS and the POSS-I plates
the maximum pairing radius is only $\sim$10'', making spurious pairings
unlikely. This is due not only to the low chance probability of another object
lying in this region near to the target but also because many chance objects which
lie so close to the target may also be deblended by the SuperCOSMOS
software Hambly et al. (2001). For reasons of poor astrometry, we have excluded all deblended
objects. The advantage of the POSS-II plates over the POSS-I plates is their better astrometry and photometry. However with the long IPHAS-POSS-I time baseline, even this better quality POSS-II astrometry cannot produce a lower minimum proper motion than using POSS-I plates and we have CCD quality photometry available from the IPHAS survey. The SuperCOSMOS scans of POSS-I plates only extend to $\delta \sim 2.5^{\circ}$, south of this we use SuperCOMSOS UK Schmidt Telescope $R$ plates.

Before beginning the proper motion survey it is important to have both  surveys on the same astrometric framework as our initial calculations will be based on the global astrometric frameworks of both surveys. IPHAS is tied to the 2MASS astrometric framework so we converted the POSS-I astrometry to the 2MASS astrometric reference frame. This was done in an identical way to the transformation of UKST $I$ plates to the 2MASS system described in Section 2.1 of Deacon \& Hambly (2007). 

In order to estimate the minimum significant positional shift that can be detected we robustly
calculated the positional errors between the POSS-I plates and the IPHAS
survey (the error estimates calculated between the IPHAS and UVEX surveys found in Figure~\ref{uvexErrplot} were calculated in the same way). This was done by identifying the same objects in each epoch and
calculating the positional differences. These were then binned by magnitude
and the error calculated (Figure \ref{Errplot}). After examining this plot, we determined the 5$\sigma$ positional shift to be at one arcsecond, this was then used as our minimum positional shift. This means our minimum proper motion will be roughly 20 mas/yr. As we will calculate relative astrometric solutions for each object in our final catalogue this number may vary slightly.

 \begin{figure}
 \setlength{\unitlength}{1mm}
 \begin{picture}(75,120)
 \includegraphics{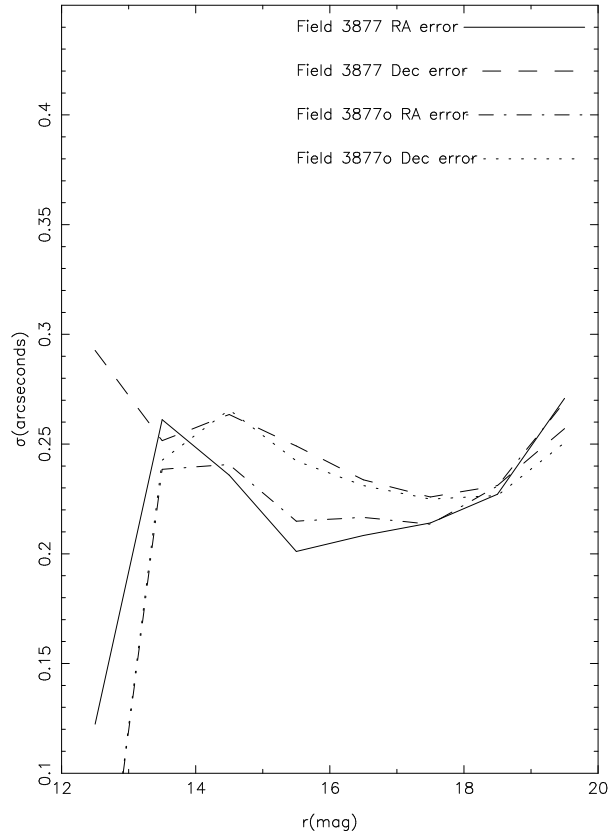}
 \end{picture}
 \caption[]{The astrometric errors (in arcseconds) between the IPHAS and POSS-I data.}
 \label{Errplot}
 \end{figure}

Our search methodology was as follows. Objects which were flagged as stellar
sources or probable stellar sources (classification flags -1 and -2, Drew et al., 2005) in IPHAS
were selected. This excludes saturated sources and hence introduces a bright limit to our survey at approximately $r$=13.5. One initial problem encountered was the difference in the sizes
of the Point Spread Functions of the two surveys. Often two stars with a small
separation which are resolved in IPHAS will be blended together on the lower resolution POSS plates leading to the erroneous conclusion that
one or both of them have moved. To remove this potential source of contamination any IPHAS object of brightness $r=x$ (where $x$ is in magnitudes) which had
another IPHAS object brighter than $x-2$ within 6 arcseconds was excluded. This magnitude difference was selected as typically objects which are two or more magnitudes fainter than an object will not significantly affect its astrometry. This pairing radius increased at brighter magnitudes, in line with the rough size of the POSS-I PSF (up to 20'' for stars brighter than $r$=9).

Subsequently, IPHAS objects which were not affected by such crowding had their
positions compared with the POSS-I data to see if they had a companion within
 an arcsecond. If they did they
were judged not to have a significant proper motion and hence were
excluded. Any potential POSS-I pair for these unpaired objects was then
searched for. First the region within 6'' was searched and if no pair was
found the region within $r_{max}= \mu_{max} \Delta t$ was searched. Any
potential pair had to have a POSS-I $R_F$ magnitude within 3$\sigma$ (where
$\sigma$ is approximated from the values for measurement errors quoted in
Hambly et al. 2001, roughly 0.2 magnitudes at best\footnote{Note as the IPHAS photometric errors are typically much smaller than POSS-I errors we ignore them in our error estimation.}) of the IPHAS $r$ magnitude and had to be stellar sources which had not been
deblended and were not in close proximity to bright stars (note this 3$\sigma$ cut could exclude high proper motion variables). To ensure that the paired POSS-I object does not have an IPHAS counterpart, the POSS-I positions were crosschecked with IPHAS positions and any object with an IPHAS pair within one arcsecond was excluded.
\subsection{Calculation of astrometric solution}
In order to gain an insight into the local astrometric accuracy of each proper motion measurement, a local relative astrometry mapping was carried out for each candidate. To do this all objects in the same IPHAS field as the target with brightnesses within one magnitude of the star in question were selected. These were then used to produce a 6 parameter plate-plate fit using SlaLib routines (Wallace, 1998) to determine the astrometric differences between the two reference frames and to estimate the random errors remaining once these differences have been corrected for. This fit was then applied and used to calculate a proper motion relative to this reference frame. This also yielded measures of the positional errors for each field. However in cases with few reference stars (i.e. $<$20) the error will be underestimated. To correct for this we carried out a series of simulations. Sets of reference stars on two different reference frames were created. These were given small random bulk offsets between the reference frames as well as individual random Gaussian errors. A fit between the reference frames was carried out and the calculated positional error compared to the indiviual positional errors used. It was found that for few reference stars the error was underestimated. We find that the correction factor is well fitted by the equation,
\begin{equation}
\frac{\sigma_{true}}{\sigma_{measured}} \approx 1 + \frac{19.8}{n_{ref}^{1.5}}
\end{equation}
Where $\sigma_{true}$ is the actual error, $\sigma_{measured}$ the measured error and $n_{ref}$ the number of reference stars. We find this relation holds fairly well down to as few as six reference stars. This correction factor was used to ensure all our quoted errors are accurate.
Where there were not enough reference stars for any fit an error calculated from the global positional error estimates shown in Figure~\ref{Errplot} was used. 
\section{Results}
The final catalogue consists of 103058 objects spread across 14126 IPHAS fields (including overlap fields) where the area of each field is roughly 0.3 sq.deg. These objects all have proper motions more significant than 5$\sigma$ where the proper motion errors were typically below 10 milliarcseconds per year (i.e. $\mu_{min} <$ 40mas). A full list of all these objects will be provided in the electronic edition. To check the sample a reduced proper motion diagram (Luyten, 1918, credited to Hertzprung) was produced. Reduced proper motion takes observables (proper motion and apparent magnitude) and combines them in such a way that the result only depends on characteristics of the star (tangential velocity and absolute magnitude). The definition we use is given below. 
   \begin{figure}
 \setlength{\unitlength}{1mm}
 \begin{picture}(75,120)
 \includegraphics{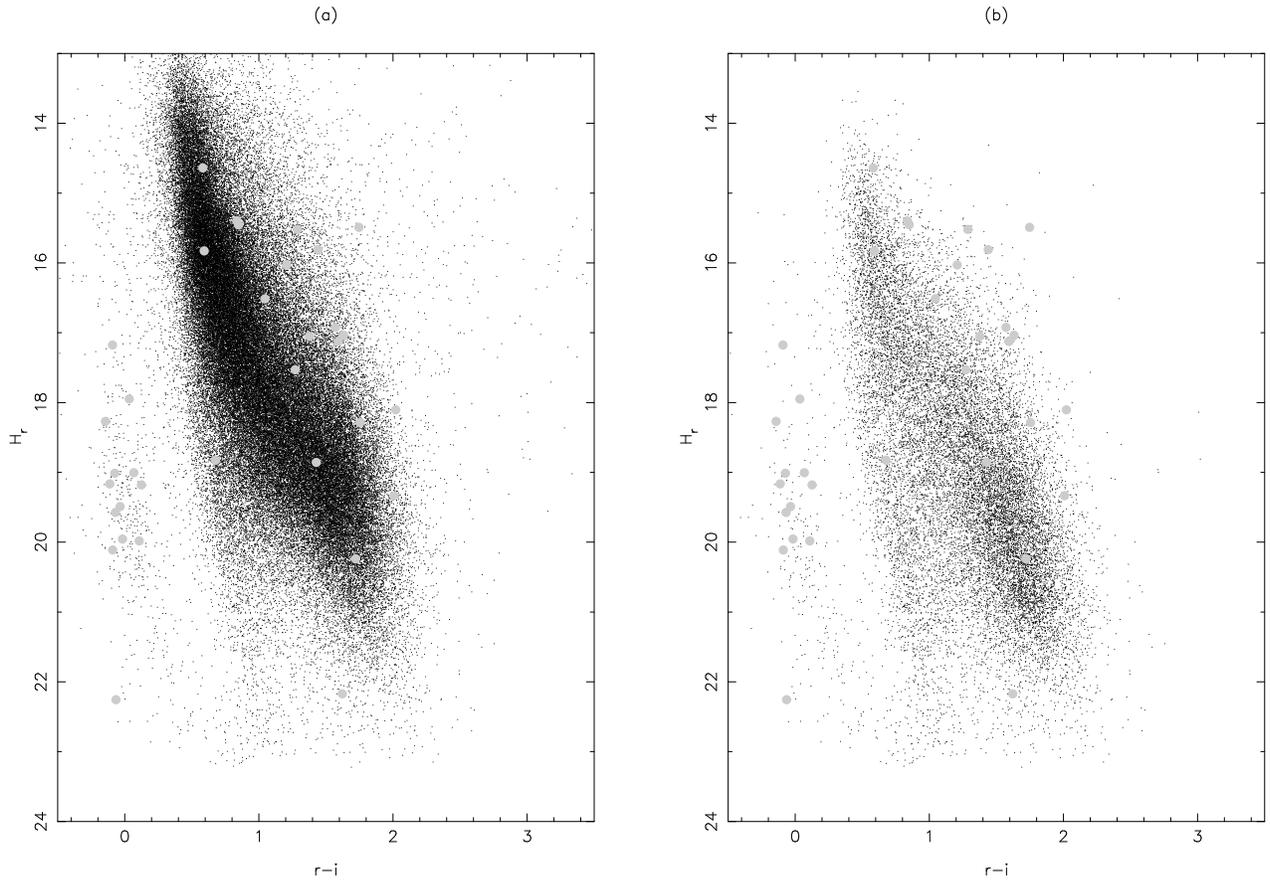}
 \end{picture}
 \caption[]{Two reduced proper motion diagrams for our dataset. (a) shows all objects in our sample, (b) shows only those with $\mu$ greater than 50mas/yr. The populations shown are as follows, the main locus is the main sequence, below and to the left are the higher velocity and bluer subdwarfs and to the left of them are the intrinsically fainter white dwarfs. The large grey dots represent the objects common between this catalogue and the catalogue of H$_\alpha$ emitters from Witham et al. (2008). These are all plotted on both panels, regardless of their proper motions.}
 \label{RPMplot}
 \end{figure}   
\begin{figure}
 \setlength{\unitlength}{1mm}
 \begin{picture}(75,65)
 \includegraphics{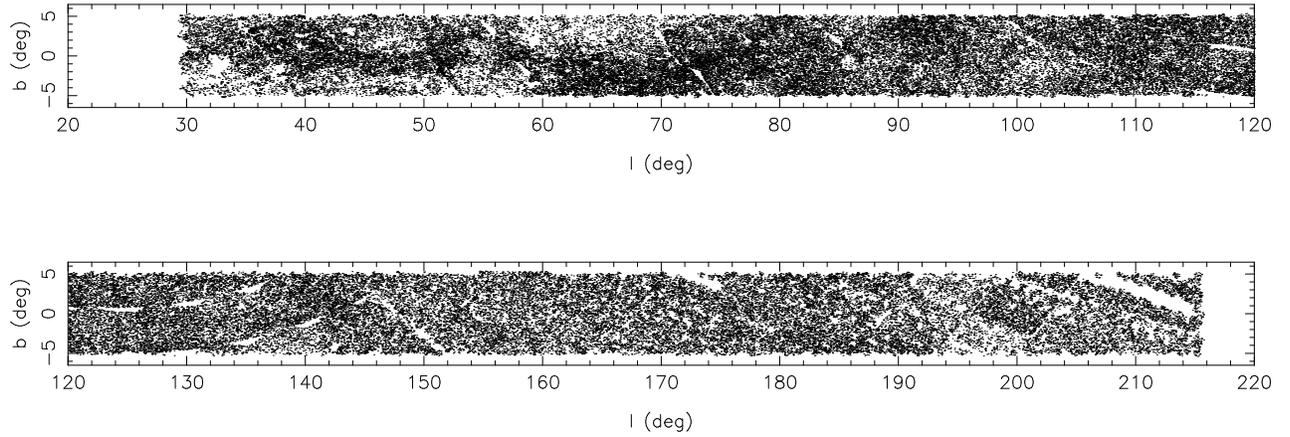}
 \end{picture}
 \caption[]{Distribution of objects in our sample across the Galactic plane. The coverage is in general good, however the coverage appears patchy in parts, particularly at low Galactic longditude ($l<$90). Also note the areas with no coverage, these either lie outside the survey area (too far south) or have not yet been covered in the survey.}
 \label{areaplot}
 \end{figure}

\begin{figure}
 \setlength{\unitlength}{1mm}
 \begin{picture}(75,65)
 \includegraphics{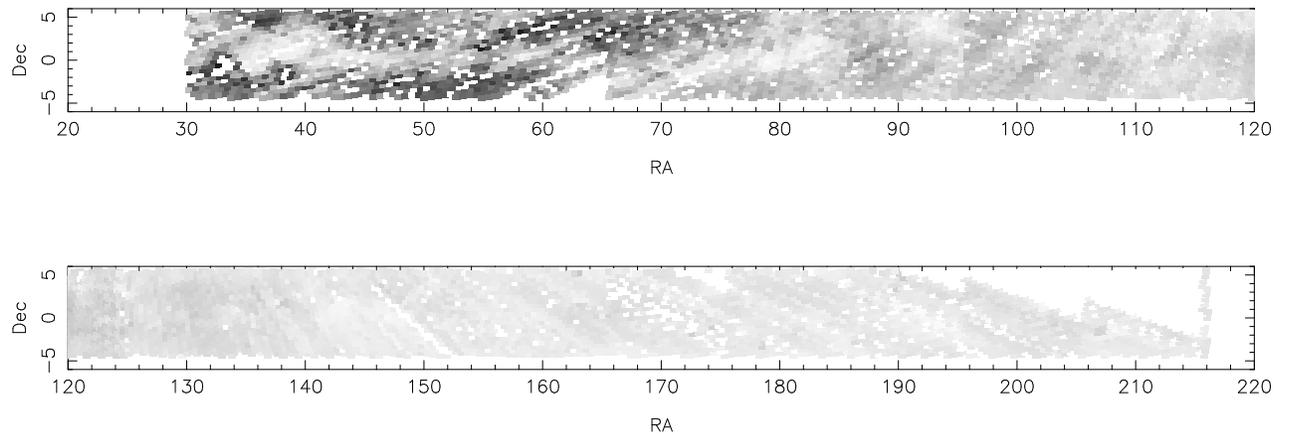}
 \end{picture}
 \caption[]{The density of stellar sources in the IPHAS survey with black being most dense and white being less dense. The larger stellar density closer to the Galactic centre along with the patches of extinction close to the plane in this region can be clearly seen.}
 \label{denseplot}
 \end{figure}
\begin{figure}
 \setlength{\unitlength}{1mm}
 \begin{picture}(75,70)
 \includegraphics{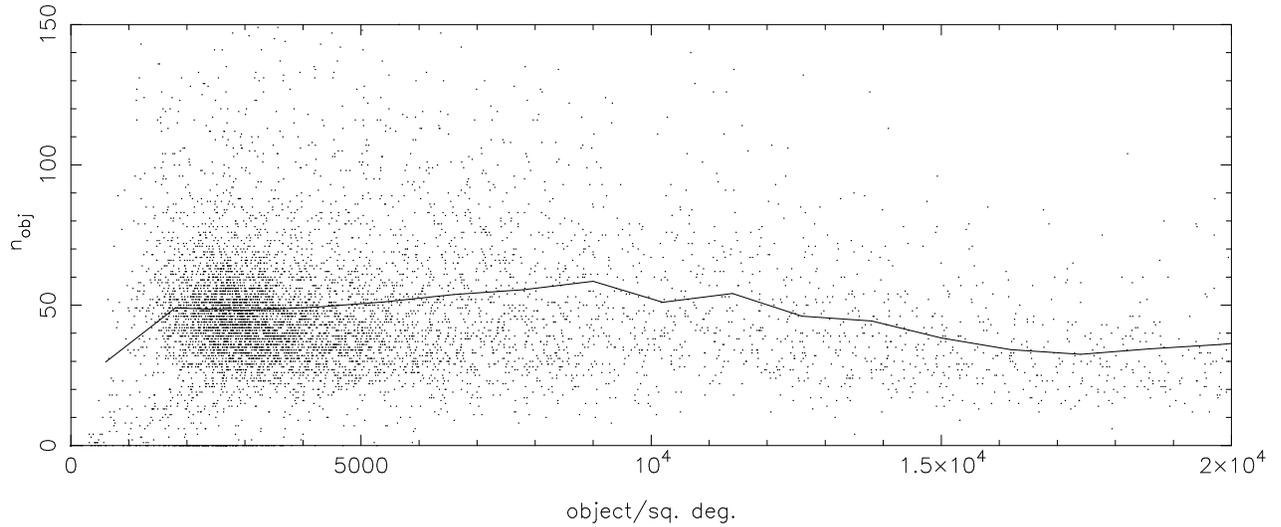}
 \end{picture}
 \caption[]{The density of stellar sources in the IPHAS survey for each field vs. the number of proper motion objects detected in each field. The solid line shows the mean number of objects for fields binned by stellar density. Note the general trend, dense fields have fewer detected objects. This is because the crowding confusion reduction algorithm removes more of the area of crowded fields.}
 \label{densecomp}
 \end{figure}
\begin{figure}
 \setlength{\unitlength}{1mm}
 \begin{picture}(75,120)
 \includegraphics{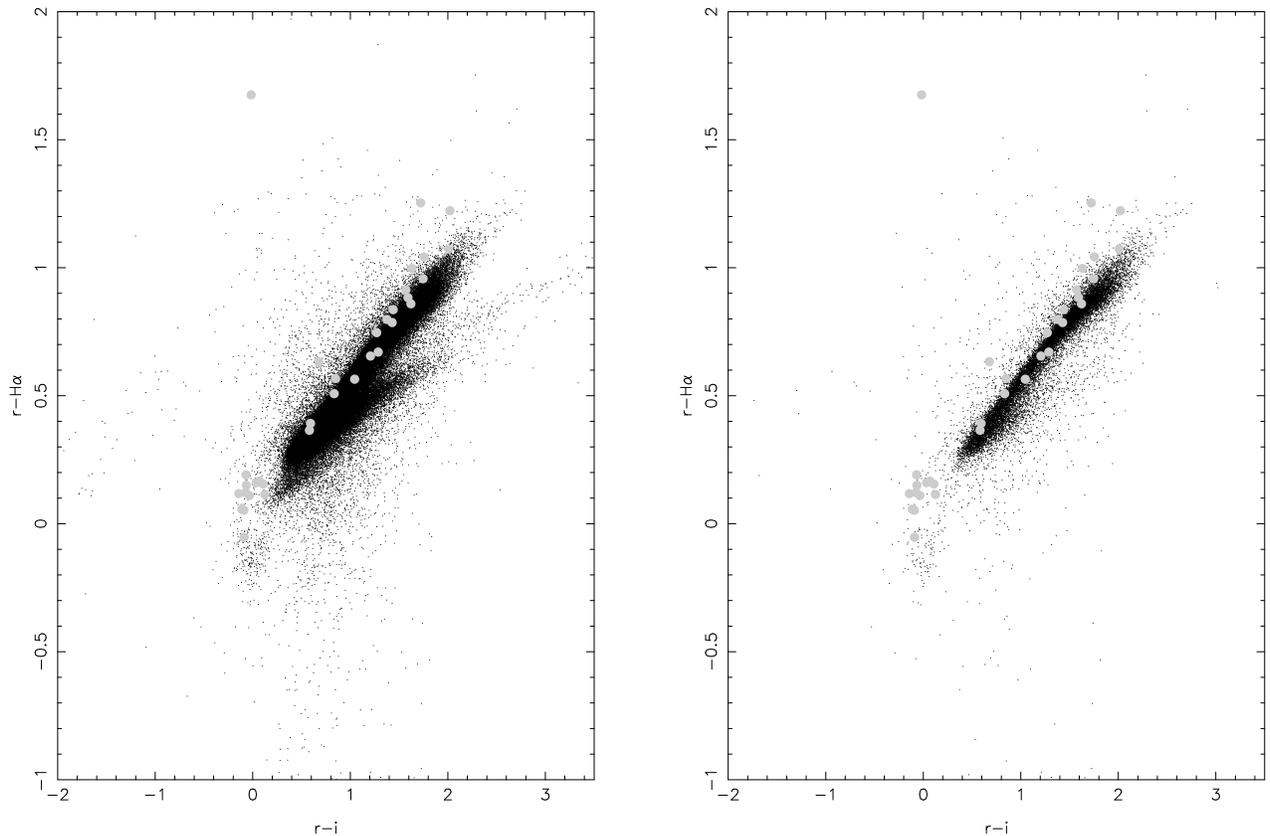}
 \end{picture}
 \caption[]{Colour-colour diagrams for the objects. The panel on the left (a) shows all the objects in our sample while the panel on the right (b) shows only those with proper motions greater than 50mas/yr. The main stellar locus runs from (0.0,0.1) to (2.0,1.0), this is a near-perfect unreddened main sequence (see Drew et al., 2005). Below and to the left lie the bluer white dwarfs and above and to the left lie potential H$_{\alpha}$ emitters. The large grey dots represent the objects common between this catalogue and the catalogue of H$_\alpha$ emitters from Witham et al. (2008). These are all plotted on both panels, regardless of their proper motions.}
 \label{ccplot}
 \end{figure}
\begin{figure}
 \setlength{\unitlength}{1mm}
 \begin{picture}(75,65)
 \includegraphics{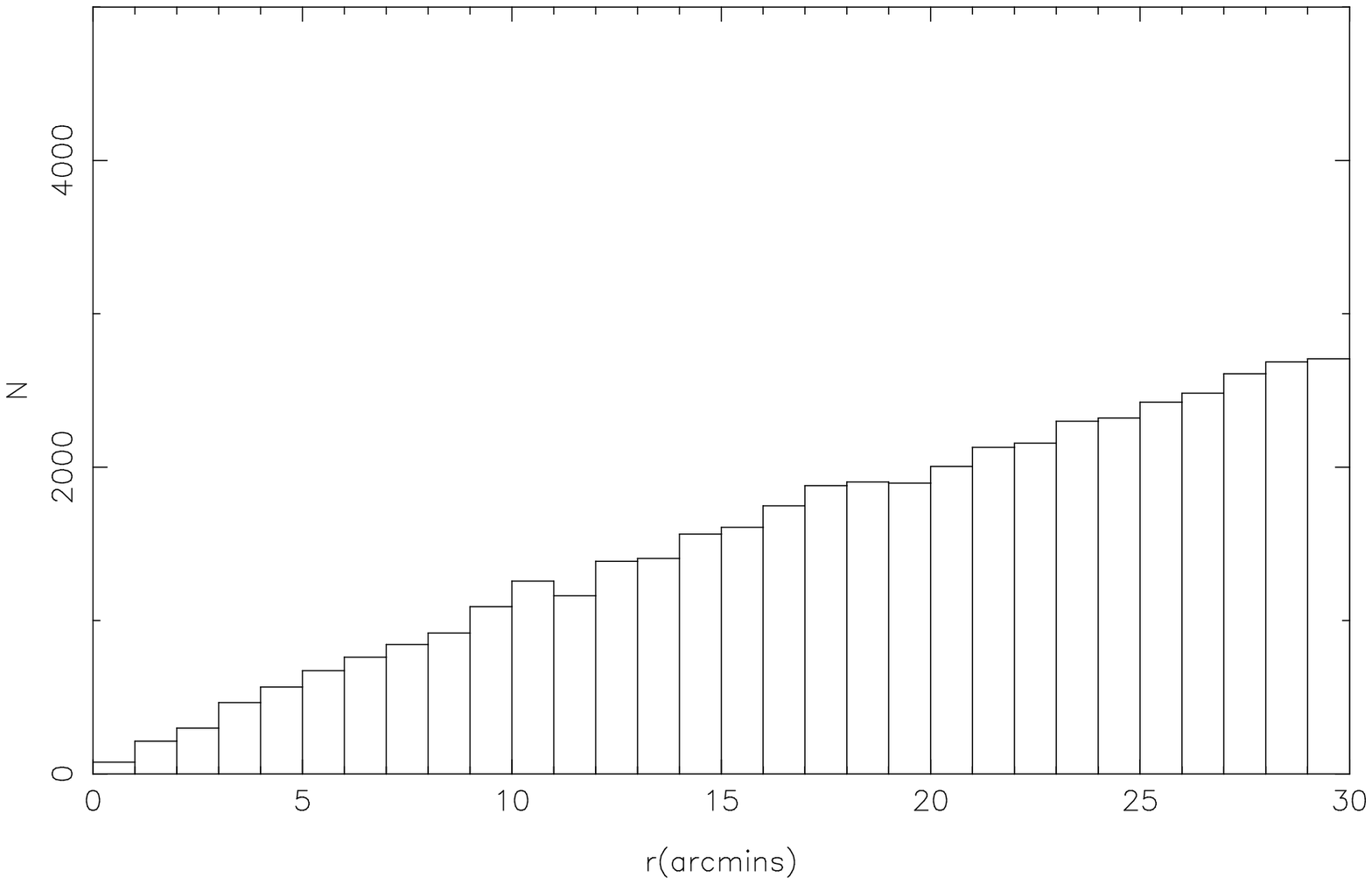}
 \end{picture}
 \caption[]{A histogram of the separations of common proper motion pairs in our sample. The trend for coincidence objects would be N$\propto r$. As we see no deviation from this trend at small seperations, we conclude that there is no significant population of true common proper motion binaries in our sample.}
 \label{sepplot}
 \end{figure}
 \begin{equation}
\begin{array}{l}
 H_r = r + 5\log_{10}\mu + 5 \log_{10}(47.4)\\
H_r = M_r + 5\log_{10}d - 5 + 5\log_{10}v_T - 5 \log_{10}(4.74) - 5\log_{10}d + 8.379\\
H_r = M_r +5\log_{10}v_T\\
\end{array}
 \end{equation}
Where $\mu$ is the proper motion in arcseconds per year, $d$ is the distance in parsecs and $v_T$ is the tangental velocity in km/s. The above definition of reduced proper motion is not the most commonly used but is useful as it removes the constants needed to convert between units. Our reduced proper motion diagram is shown in Figure~\ref{RPMplot}. The form
is roughly what we would expect from a standard Galactic stellar population
with clearly identifiable dwarf, subdwarf and white dwarf loci. However after
we studied the spatial distribution of objects it was found that there were
several fields with many (more than 250) objects. After some investigation it
became clear that these fields had poor astrometric solutions in the IPHAS
data (mostly due to poor observing conditions). When we examined a histogram of number of detected objects per field it was found that these fields lay beyond the point where the main distribution had died away. Additionally when the reduced proper motion diagrams for objects in these fields was examined it was found that it did not contain the expected population distributions, indicating that the proper motion determinations were not correct. Hence any object lying in these fields was excluded from the final catalogue. A plot of the spatial distribution of the remaining objects can be found in Figure~\ref{areaplot}. It shows that for the majority of the northern plane, the coverage is good with a few patches of incompleteness. However moving along the plane towards the Galactic centre the number of objects drops off dramatically. This is due to our selection criteria excluding crowded regions as well as large numbers of objects in these regions being blended with other images (again a result of high stellar density). This can be seen in Figure~\ref{denseplot} which shows the density of stellar sources in each IPHAS field: there are clearly fewer high proper motion objects detected in areas of higher stellar density~\footnote{The general trend towards more crowded fields towards the Galactic centre can be seen in Figure 3 of Gonzalez-Solares et al. 2008}. This is also shown by the inverse correleation between the density of stellar sources in a field and the typical number of detected proper motion sources in that field (see Figure~\ref{densecomp}). Figure~\ref{ccplot} shows an IPHAS colour-colour plot for our objects. The main locus is a clear, unreddened main sequence (see Drew et al., 2005), widened by the fact that the IPHAS photometry is not yet globally calibrated. Approximately 96\% of objects in the catalogue lie on or close to this main sequence. There is also a white dwarf locus present lying below and to the left of the main sequence. Many objects lie above and to the left of the main sequence. While this may suggest H$\alpha$ emission, it may also be due to poor photometry in a particular field. Hence rather than select all these as potential H$\alpha$ emitters, in the next section we will use the study of Witham et al. (2008) to identify objects which appear to have significant H$\alpha$ emission relative to the main sequence on the particular field . Finally we checked our sample for common proper motion binaries. To investigate if we had a distinct population of common proper motion binaries, we plotted a histogram of the separations of all the objects with proper motions within 2$\sigma$ of each other. The trend for coincidence objects should be N$\propto r$ and any excess above this at small separations would indicate a separate population of physically bound common proper motion objects. Figure~\ref{sepplot} shows our histogram, clearly there is no distinct population of common proper motion binaries present.
\subsection{Comparisons with other IPHAS studies}
As stated earlier, the IPHAS survey is currently being exploited for many different scientific goals. One study utilising IPHAS photometry is that of Witham et al. (2008). Here IPHAS photometry is used to identify objects which lie significantly above the main stellar locus on a colour-colour diagram similar to Figure~\ref{ccplot}. As there will be offsets in the photometry from field to field, Witham et al. (2008) identifies potential H$\alpha$ emitters relative to the colour-colour diagram for the field the object lies in. Hence objects which appear to be  H$\alpha$ emitters due to the poor photometry of an individual field are not included in Witham et al's sample. This allows us to treat this dataset as a clean sample of potential H$\alpha$ emitters. Cross-referencing this with our own proper motion sample will remove highly reddened (and distant) Be stars from the Witham sample and should leave only potential Cataclysmic Variables candidates, dMe stars and non-DA white dwarfs (ie. nearby stellar sources showing either $H_{\alpha}$ emission or less than expected $H_{\alpha}$ absorption). In this cross-referencing, we also included objects found in our study with proper motions between 0.2 and 0.15 arcseconds per year and objects with $r$ magnitudes between 19 and 20. These were not included in the final catalogue as these objects were found to suffer from a high level of contamination. 

The thirty six crossmatches are shown in Table~\ref{Halpha}. Note eight crossmatches were excluded from this list and from Figures~\ref{RPMplot} and~\ref{ccplot} after inspection of the images by eye found that they may be blended objects.
\begin{table*}
 \begin{minipage}{170mm}
  \caption{Objects common between our catalogue and the H$_{\alpha}$ catalogue of Witham et al. (2008). IPHASJ225040+632838 is the known proper motion CV system GD 552 (Greenstein \& Giclas 1978), IPHASJ043839+410931 is (GD 61 Giclas, Burham \& Thomas, 1965), IPHASJ210951+425705 is EGGR 334 (Greenstein, 1974) and IPHASJ032825+580645 is the known high proper motion star LSPM J0328+5806 (Lepine \& Shara, 2005). spectral type sources $^1$ WHT spectroscopy, $^2$ FAST spectroscopy, $3$ Giclas, Burham \& Thomas (1965), $^4$ Greenstein (1974), $^5$ Greenstein \& Giclas 1978}
\label{Halpha}
\footnotesize
  \begin{tabular}{lccrcccccc}
  \hline
Name&Position&$\mu_{\alpha}$&$\mu_{\delta}$&$\sigma_{\mu_{\alpha}}$&$\sigma_{\mu_{\delta}}$&$r$&$i$&$H_{\alpha}$&SpT\\
&&''/yr&''/yr&''/yr&''/yr&&&\\
  \hline
IPHASJ000528+663951&00 05 28.05 +66 39 51.5&0.031&-0.051&0.009&0.009$^2$&13.582&13.001&13.190&\\ 
IPHASJ002156+630635&00 21 56.62 +63 06 35.8&-0.044&-0.026&0.005&0.005$^2$&17.105&17.183&16.993&\\ 
IPHASJ010749+582709&01 07 49.39 +58 27 09.3&0.022&-0.031&0.005&0.006$^2$&14.171&13.279&13.657&\\ 
IPHASJ031119+600110&03 11 19.21 +60 01 10.8&0.044&-0.053&0.005&0.006$^1$&18.181&15.995&17.058&\\
IPHASJ032327+534705&03 23 27.39 +53 47 05.4&0.061&-0.054&0.005&0.005$^1$&16.734&16.980&16.576&\\ 
IPHASJ032825+580645&03 28 25.12 +58 06 45.8&0.144&-0.042&0.006&0.007$^2$&17.684&17.917&17.550\\ 
IPHASJ032905+563606&03 29 05.01 +56 36 06.8&-0.019&-0.034&0.006&0.007$^2$&14.704&13.503&14.053&\\ 
IPHASJ033805+563518&03 38 05.68 +56 35 18.7&0.029&-0.038&0.006&0.006$^2$&13.743&12.461&13.077&dMe$^2$\\
IPHASJ034042+573053&03 40 42.96 +57 30 53.7&0.069&-0.033&0.006&0.006$^2$&13.724&12.685&13.164&dM$^2$\\ 
IPHASJ040147+540650&04 01 47.07 +54 06 50.8&0.044&-0.058&0.006&0.007$^2$&16.122&14.727&15.361&\\
IPHASJ043839+410931&04 38 39.38 +41 09 31.9&-0.011&-0.110&0.005&0.006$^2$&14.673&14.816&14.556&DB$^3$\\
IPHASJ045400+470031&04 54 00.68 +47 00 31.0&0.004&-0.023&0.006&0.003$^1$&17.723&17.691&17.563&\\
IPHASJ053015+251137&05 30 15.51 +25 11 37.3&0.031&0.044&0.005&0.005$^2$&13.429&12.580&12.862&dM$^2$\\ 
IPHASJ055551+324150&05 55 51.14 +32 41 50.3&-0.037&-0.001&0.006&0.006$^2$&17.829&17.707&17.599&DC$^2$\\ 
IPHASJ055752+274641&05 57 52.90 +27 46 41.8&0.025&-0.044&0.005&0.006$^1$&17.264&17.415&17.189&DC$^2$\\ 
IPHASJ061409+171136&06 14 09.36 +17 11 36.0&0.003&-0.044&0.006&0.006$^2$&16.670&14.917&15.628&\\
IPHASJ062809+163158&06 28 09.40 +16 31 58.7&-0.046&-0.012&0.006&0.006$^2$&17.833&17.909&17.642&DB$^1$\\ 
IPHASJ183523+014245&18 35 23.26 +01 42 45.4&0.154&0.015&0.006&0.006$^2$&17.700&16.172&16.921&\\
IPHASJ184306+004111&18 43 06.88 +00 41 11.3&-0.031&-0.048&0.005&0.005$^2$&18.020&18.071&17.930&\\ 
IPHASJ185929-040304&18 59 29.38 $-$04 03 04.3&0.051&-0.027&0.005&0.005$^2$&13.335&11.604&12.384&\\ 
IPHASJ190132+145807&19 01 32.77 +14 58 07.6&0.082&0.076&0.006&0.007$^1$&15.905&15.870&15.823&DC$^2$\\
IPHASJ190142-043621&19 01 42.09 $-$04 36 21.1&0.001&-0.034&0.007&0.006$^1$&15.999&14.622&15.201&\\
IPHASJ190338-025232&19 03 38.54 $-$02 52 32.4&0.032&-0.010&0.005&0.005$^1$&16.514&15.243&15.768&\\
IPHASJ191733+031937&19 17 33.35 +03 19 37.9&0.138&-0.032&0.005&0.006$^2$&15.406&14.984&14.830\\\
IPHASJ192206+053238&19 22 06.11 +05 32 38.5&0.031&0.000&0.005&0.005$^2$&16.254&14.723&15.450&\\
IPHASJ201409+265254&20 14 09.92 +26 52 54.1&0.035&0.038&0.005&0.006$^1$&15.113&13.408&14.187\\
IPHASJ210541+534334&21 05 41.78 +53 43 34.5&0.025&-0.020&0.005&0.005$^2$&14.891&13.451&14.055&\\
IPHASJ210923+515607&21 09 23.85 +51 56 07.8&0.027&0.038&0.006&0.006$^2$&17.632&15.631&16.531&\\
IPHASJ210951+425705&21 09 51.24 +42 57 05.1&0.192&-0.018&0.010&0.005$^1$&15.552&15.559&15.340&DB$^4$\\
IPHASJ215029+554250&21 50 29.23 +55 42 50.6&0.027&0.006&0.005&0.005$^1$&17.477&15.492&16.283&\\
IPHASJ223541+590745&22 35 41.31 +59 07 45.7&0.026&-0.005&0.004&0.005$^1$&16.644&16.729&16.547&\\
IPHASJ224918+614903&22 49 18.57 +61 49 03.9&-0.039&-0.023&0.004&0.005$^2$&17.508&17.475&17.436&DA$^1$\\
IPHASJ225040+632838&22 50 40.03 +63 28 38.2&0.102&-0.037&0.005&0.006$^2$&16.389&16.406&14.714&CV$^5$\\
IPHASJ232003+571736&23 20 03.28 +57 17 36.6&-0.035&0.000&0.007&0.007$^2$&13.537&12.965&13.170&\\
IPHASJ232158+581034&23 21 58.03 +58 10 34.6&0.030&-0.011&0.006&0.005$^1$&16.037&14.454&15.129&\\
IPHASJ232908+615911&23 29 08.78 +61 59 11.0&-0.183&-0.067&0.007&0.004$^2$&17.626&17.304&17.330&DC$^1$\\
  \hline
\normalsize
\end{tabular}
\end{minipage}
\end{table*}
Examining Figure~\ref{ccplot} we can see that many of the grey dots (representing Witham et al.'s H$\alpha$ emitters with significant proper motions) fall along the main sequence. It is possible that these are true H$\alpha$ emitters and appear in this part of the diagram due to uncorrected field to field photometric offsets or some selection effect. Of these objects one (IPHASJ053015+251137) appears to share a common proper motion with the nearby (separation 42'') star TYC 1852-777-1 (Hog et al., 1998). The two proper motions agree within one sigma implying these are a true bound pair or part of the same moving group. Three other objects redder than $r-i=0.4$ have spectra from FAST follow-up observations of IPHAS sources. Of these  one (IPHASJ033805+563518) was found to be an M dwarf with H$\alpha$ emission. The question remains as to why these objects appeared in Witham et al.'s catalogue. Witham et al. fitted a curve to the unreddened main sequence in each field and identified emitters as objects which lay significantly above this curve. The two non-emitting M dwarfs lie in the brightest selection bin of Witham et al.'s selection process ($r <$16). Here the number of objects defining the unreddened main sequence will be smallest. It could be that these are marginal selections where the unreddened main sequence is affected by poor statistics. It is also possible that these are objects with variable weak H$\alpha$ emission. One moderately red object (IPHASJ191733+031937) appears to lie on the subdwarf sequence.

Fifteen of the cross matches objects which appear to lie on the white dwarf sequence in the reduced proper motion diagram (Figure~\ref{RPMplot}). Of these IPHASJ225040+632838 is the low state CV system GD 552 (Greenstein \& Giclas 1978). Another two, IPHASJ043839+410931 (GD 61, Giclas, Burham \& Thomas, 1965) and IPHASJ210951+425705 (EGGR 334, Greenstein, 1974) are known DB white dwarfs. Additionally three objects had spectra taken in the IPHAS spectroscopic follow-up programme with the FAST spectrograph on the 1.5m Tillinghast telescope on Mount Hopkins. Of the remaining nine objects, three had spectra taken using the ISIS spectrograph on the William Herschel Telescope (WHT) on La Palma. These spectra were used to provide rough spectral classifications which can be found in Table~\ref{Halpha}. Seven of the eight spectrally classified objects which lie bluewards of $r-i$=0.4 in the colour-colour diagram (excluding the known CV GD 552) are non-DA white dwarfs. Hence we believe the remaining objects are good non-DA white dwarf candidates.

Valdivielso et al. (2008) have produced a sample of young, low mass objects using IPHAS data. Clearly identifying the proper motions of such objects could establish a connection with a known star forming association or moving group. Unfortunately none of these objects appear in our catalogue.
\section{Discussion}
\begin{figure}
 \setlength{\unitlength}{1mm}
 \begin{picture}(75,105)
 \includegraphics{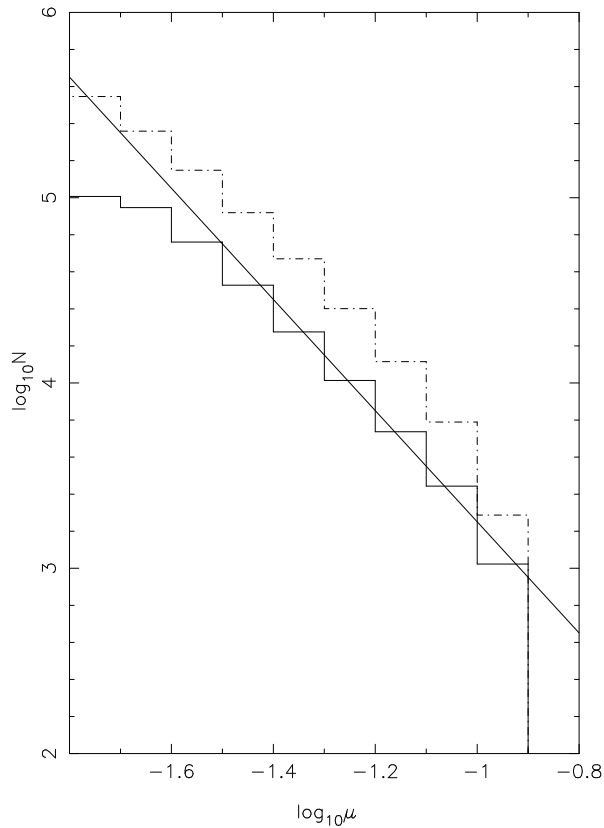}
 \end{picture}
 \caption[]{A cumulative proper motion histogram for our the objects in our catalogue. The solid line represents the N$\propto \mu^{-3}$ relation that would be expected with no incompleteness. It is clear that our survey begins to become incomplete below about 60 milliarcseconds per year and that below 20 millarcseconds per year there are virtually no objects. The dotted line represents the study of Gould \& Kollmeier (2004). Clearly their study is complete to lower proper motions than ours. In the region where our survey is most complete there is a factor of 2 difference between the numbers.}
 \label{PMhist}
 \end{figure}

In order to provide a rough estimate of our completeness, we plotted a cumulative proper motion histogram. This is shown in Figure~\ref{PMhist}. Assuming uniform spatial and velocity distributions and a fully complete survey, the distribution should scale as N$\propto \mu^{-3}$. This is represented by the solid line in the plot. It is clear that we begin to become incomplete below 60 milliarcseconds per year. This is due to a combination of our limiting magnitude and some objects falling in fields with poor astrometry (hence having proper motions which are not significant enough). Below about 25 mas/yr it is clear the distribution flattens off and we can say we have no significant population below this mark. We have also compared our results with those in Gould \& Kollmeier (2004). Figure~\ref{PMhist} shows that in the proper motion range where both surveys have similar proper motion completeness, we have half the number of objects that Gould \& Kollmeier have. This is despite the two surveys having similar areas (both around 1400 sq. deg.). However our survey covers a much more crowded area than theirs. Deacon, Hambly \& Cooke (2005) calculated the area lost to bright and blended stars across the southern sky. Examining their Figure 10, it is clear that in the southern regions of the sky at similar Galactic latitude to ours, the completeness is often 50\% or worse. Hence we believe this difference in numbers is due the more crowded nature of our survey area.  

The IPHAS survey consists of 15270 pointings, which between them cover the 1800 square-degree survey area twice or more. Hence simply taking the size of the detector and multiplying it by the number of fields our survey covers (12362) will not yield an accurate estimate of our current survey area. A rough estimate can be provided by multiplying the fraction of the fields we cover (approximately 81\%) by the total final survey area of 1800 sq. deg. This yields and approximate area for our proper motion survey of 1457 sq. deg. However as stated above, due to crowding we are only likely to identify proper motion objects in roughly half this total area. Once data from the few unobserved IPHAS fields have been relaesed we will apply the same method to the remaining fields, completing our proper motion survey.

In calculating our astrometric solutions we use sets of reference stars. These may have small bulk motions. Additionally for the objects where we have too few reference stars the raw IPHAS positions are used. These are tied to the 2MASS (Skrutskie et al. 2006) system using reference stars. Hence we will measure proper motions relative to these reference stars rather than absolute proper motions. Lepine (2008) also encountered this problem. They concluded that the difference between absolute and relative proper motions was typically less than their measurement errors. As our measurement errors are similar to theirs (typicall below their quoted global errors of 8mas/yr in each axis), we deduce that any offset between the relative and absolute proper motions of our sample will also be below our calculated errors. 
\section{Conclusions}
We have completed the first comprehensive wide field proper motion survey of the northern Galactic plane ($|b| < 5^{\circ}$) covering proper motions between 150 and approximately 30 arcseconds per year. This sample covers a large section (1457 sq. deg.) of the northern plane and contains 57249 objects with significant proper motions. We also identify seventeen objects in common between our catalogue and the H$\alpha$ emission catalogue of Witham et al. (2008). These objects fell in to two distinct groups, a blue group dominated by non-DA white dwarfs and a red group dominated by maginally selected ordinary main sequence objects. This sample will clearly be useful in the study of populations such as white dwarfs and subdwarfs in the Galactic plane. We will seek to complete the catalogue for the full survey area and will use the upcoming UVEX data to extend it to higher proper motions above the current imposed limit of 0.15 arcseconds per year. 
\section*{Acknowledgments}
This paper uses data from the SuperCOSMOS Sky Survey and from the INT Photometric H$\alpha$ Survey of the northern Galactic plane (IPHAS) carried out at the Isaac Newton Telescope (INT). The INT is operated on the island of La Palma by the Isaac Newton Group in the Spanish Observatorio del Roque de los Muchachos of the Instituto de Astrofisica de Canarias. All IPHAS data are processed by the Cambridge Astronomical Survey Unit, at the Institute of Astronomy in Cambridge.. N.R.D. is funded by NOVA and by NWO-VIDI grant 639.041.405 to Paul Groot. DS acknowledges a STFC Advanced Fellowship. This paper makes use of Slalib routines (see Wallace, 1998). This research has made use of the SIMBAD database, operated at CDS, Strasbourg, France. We thank the FAST observers for their assistance with obtaining the follow-up spectroscopy of IPHAS emitters. The 1.5m Tillinghast telescope is located near Mt.Hopkins in Arizona and operated on behalf of the Smithsonian Astrophysical Observatory. the authors would like to thank Boris Gaensicke, Christian Knigge, Quentin Parker and Stuart Sale for their helpful comments.

\appendix
\section{Example Data Table}
The data tables for this paper will be available electronically. Here we give an example of one of the data tables.
\begin{table*}[h]
 \begin{minipage}{170mm}
  \caption{An example of the data tables available electronically for this paper.$^1$ indicates astrometric solutions calculated from reference stars, $^2$ implies the astrometic errors are drawn from global error estimates. The Modified Julian date (MJD) and the position are both taken from the IPHAS observations.}
\label{example}
\footnotesize
  \begin{tabular}{lcccccccccc}
  \hline
Name&Position&$\mu_{\alpha}$&$\mu_{\delta}$&$\sigma_{\mu_{\alpha}}$&$\sigma_{\mu_{\delta}}$&$r$&$i$&H$\alpha$&MJD\\
&J2000&''/yr&''/yr&''/yr&''/yr&&&&\\
  \hline
IPHASJ000001+575210&00 00 01.48 +57 52 10.9&-0.009&-0.034&0.008&0.004$^2$&15.278&14.700&14.956&53666\\ 
IPHASJ000001+652057&00 00 01.81 +65 20 57.0&0.002&-0.029&0.007&0.005$^2$&14.995&14.214&14.627&53664\\ 
IPHASJ000005+644818&00 00 05.67 +64 48 18.2&-0.001&-0.044&0.006&0.005$^2$&18.923&17.575&18.455&53664\\ 
IPHASJ000005+602518&00 00 05.81 +60 25 18.0&0.036&0.010&0.005&0.005$^1$&17.328&16.598&16.927&53665\\ 
IPHASJ000009+652017&00 00 09.09 +65 20 17.2&-0.078&0.008&0.007&0.006$^2$&19.885&18.113&19.235&53664\\ 
IPHASJ000010+602519&00 00 10.07 +60 25 19.7&-0.059&-0.025&0.009&0.007$^2$&13.659&12.837&13.184&53666\\ 
IPHASJ000013+582109&00 00 13.26 +58 21 09.5&-0.032&-0.029&0.007&0.004$^2$&17.388&15.783&16.565&53665\\ 
IPHASJ000014+631007&00 00 14.91 +63 10 07.1&0.035&-0.004&0.006&0.006$^1$&16.513&15.365&15.879&53666\\ 
IPHASJ000017+664726&00 00 17.51 +66 47 26.4&-0.023&-0.05&0.006&0.006$^2$&12.748&12.328&12.519&54009\\ 
IPHASJ000017+620139&00 00 17.51 +62 01 39.0&0.037&-0.001&0.007&0.008$^1$&17.660&15.958&16.761&53244\\ 
IPHASJ000018+592623&00 00 18.76 +59 26 23.4&0.072&-0.027&0.004&0.007$^1$&16.140&14.542&15.308&53666\\ 
IPHASJ000023+661414&00 00 23.88 +66 14 14.3&0.038&0.005&0.005&0.007$^1$&17.770&16.107&16.977&53666\\ 
  \hline
\normalsize
\end{tabular}
\end{minipage}
\end{table*}
\label{lastpage}

\end{document}